\documentclass[aps,prb,twocolumn,superscriptaddress,showpacs,natbib,a4paper]{revtex4-1}

\usepackage[T1]{fontenc}
\usepackage[latin1]{inputenc}
\usepackage{amsfonts}
\usepackage{color}
\usepackage{dcolumn} 

\usepackage[dvips]{graphicx}
\usepackage[hypertex,breaklinks=true,colorlinks=false]{hyperref}
\usepackage[active]{srcltx}

\usepackage{ifthen}
\usepackage{ulem}

\newcommand{\journal}[4]
{\ifthenelse{\equal{#1}{pr}}{
Phys Rev. {\bf #2}, \href{http://link.aps.org/abstract/PR/v#2/e#3}{#3} (#4)}
{\ifthenelse{\equal{#1}{prl}}{
\prl {\bf #2}, \href{http://link.aps.org/abstract/PRL/v#2/e#3}{#3} (#4)}
{\ifthenelse{\equal{#1}{prb}}{
\prb {\bf #2}, \href{http://link.aps.org/abstract/PRB/v#2/e#3}{#3} (#4)}
{\ifthenelse{\equal{#1}{arxiv}}{preprint
\href{http://arxiv.org/abs/#2.#3}{arXiv:#2.#3}}
{\ifthenelse{\equal{#1}{rmp}}{
\rmp {\bf #2}, \href{http://link.aps.org/abstract/RMP/v#2/e#3}{#3} (#4)}
{\ifthenelse{\equal{#1}{cond-mat}}{preprint
\href{http://arxiv.org/abs/cond-mat/#2}{cond-mat/#2}}
{\ifthenelse{\equal{#1}{pre}}{
\pre {\bf #2}, \href{http://link.aps.org/abstract/PRE/v#2/e#3}{#3} (#4)}
{#1 {\bf #2}, #3 (#4)}}}}}}}}

\newcommand{\journaldoi}[5]{#1\ {\bf #2}, \href{http://dx.doi.org/#5}{#3} (#4)}

\begin{document}

\title{R\'enyi entropy of a line in two-dimensional Ising models}
\author{J.-M. St\'ephan}
\author{G. Misguich}
\author{V. Pasquier}
\affiliation{
Institut de Physique Th\'eorique,
CEA, IPhT, CNRS, URA 2306, F-91191 Gif-sur-Yvette, France.}

\date{June 9th, 2010}
\begin{abstract}
We consider the two-dimensional (2d) Ising model on a infinitely long cylinder and study the
probabilities $p_i$ to observe a given spin configuration $i$ along a circular section of the cylinder.
These probabilities also occur as eigenvalues of reduced density matrices in some Rokhsar-Kivelson wave-functions.
We analyze the subleading constant to the R\'enyi entropy $R_n=1/(1-n) \ln (\sum_i p_i^n)$
and discuss its scaling properties at the critical point. 
Studying three different microscopic realizations,
 we provide numerical evidence that it is universal and behaves in a step-like fashion
 as a function of $n$, with a discontinuity at the Shannon point $n=1$. 
As a consequence, a field theoretical argument based on the replica trick would fail to give the correct value at this point.
 We nevertheless compute it numerically with high precision.
Two other values of the R\'enyi parameter are of special interest: 
$n=1/2$ and $n=\infty$ are related in a simple way to the Affleck-Ludwig boundary entropies associated to free and fixed
boundary conditions respectively.
\end{abstract}
\maketitle

\section{Introduction}

\begin{figure}
\begin{center}
\includegraphics[width=6cm]{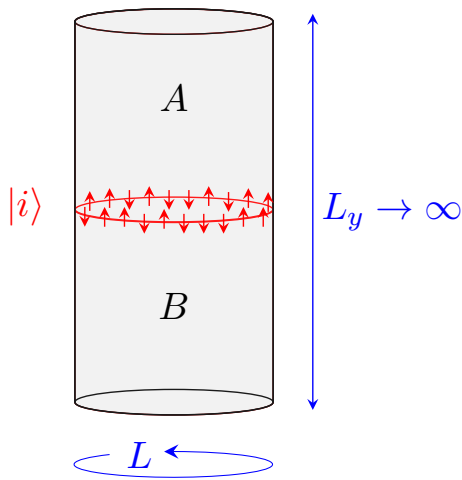}
\caption[1]{(Color online) Cylinder geometry with $L_y\gg L$.
 A probability $p_i$ is associated to each spin configuration $i$ of the boundary (red circle) between $A$ and $B$.} 
\label{fig:cyl}
\end{center}
\end{figure}

The entanglement (or Von Neumann) entropy is in general a difficult quantity to compute in two-dimensional
quantum lattice models.\cite{afov08} In Ref.~\onlinecite{fm07} it was however shown that for a particular type of wave functions,
of type dubbed ``Rokhsar-Kivelson'' (RK), and for particular geometries, the calculation simplifies considerably.
A lattice model of statistical mechanics can be used to define a Rokhsar-Kivelson wave-function
as follows:\cite{rk88,henley04}
\begin{equation}
 | {\rm RK} \rangle = \frac{1}{\sqrt \mathcal{Z}} \sum_c e^{-\frac{1}{2} E(c)}
|c\rangle.
	\label{eq:RK}
\end{equation}
where the sum runs over the classical configurations and $E(c)$ is the energy 
associated to $c$ (interactions are assumed to be short-ranged), and the normalization factor involves the classical partition function $\mathcal{Z}$.
For such a state, it has been shown in Ref.~\onlinecite{stephan09} that the eigenvalues of the reduced density matrix
of a semi-infinite cylinder
(with a finite circumference $L$, see Fig.~\ref{fig:cyl}) are simply the classical probabilities
$p_i$ to observe a given configuration $i$ at the boundary between $A$ and $B$.
In turn, these probabilities can be obtained from the dominant eigenvector of the transfer matrix of the classical model.
So, the complete entanglement spectrum is encoded in the dominant eigenvector of the classical transfer matrix.
\cite{foo1}
In this work, we concentrate on the situation where the classical model is a two-dimensional Ising model.
Each probability $p_i$ is therefore associated to a given configuration $i$
of the spins along the ``ring'' of length $L$ which separates the regions $A$ and $B$ (Fig.~\ref{fig:cyl}).
Specifically, we are interested in the behavior of the R\'enyi entropies
\begin{equation}
R_{n>0}=\frac{\ln(Z_n)}{1-n}\;\;\; Z_n=\sum_i p_i^n,
\end{equation}
including its limit
\begin{equation}
\lim_{n\to 1} R_n=-\sum_i p_i\ln(p_i),
\end{equation}
which is the Shannon entropy (or Von Neumann in the quantum/RK point of view\cite{stephan09}).

As discussed in previous studies,\cite{fm06,hsu09,stephan09}
$R_n(T,L)$ scales linearly with perimeter $L$ of the cylinder, even at the critical temperature.
 However, the most interesting piece of information is the first {\it subleading} correction, $r_n(T)$.
For a given temperature $T$, the later is defined through an expansion of $R_n(T,L)$ for large $L$:
\begin{equation}
R_n(T,L)\simeq a_n(T) L+r_n(T)+o(1)
\end{equation}
and is of order one.\cite{foo2}
Contrary to the coefficient $a_n$,  $r_n$ have been argued to be universal.
In the case of Ising models, $r_1(T>T_c)=0$ in the high temperature phase and $r_1(T<T_c)=\ln(2)$ in the low temperature phase.\cite{stephan09}
At the critical point, the previous numerical calculations (up to $L=36$) lead to
$r_1(T=T_c)\simeq 0.2544$.\cite{foo3}
The numerical results presented in Sec.~\ref{sec:2dTM} significantly increase
the precision on this number: $r_1(T_c)=0.2543925(5)$.
Furthermore, we confirm its universal character by checking the agreement between 
three microscopically different realizations of the 2d critical Ising models: on the square and triangular lattice,
and using the Ising chain in transverse field (ICTF).
At present, we are not aware of any field theory method which is able to compute this number.

In Sec.~\ref{sec:mu} we analyze the finite-size scaling  of $r_1(\mu,L)$ in the vicinity of the critical point, using numerical
(but exact) calculations
for the ICTF. There, the parameter $\mu$ measures the ratio of the spin-spin interaction over the strength of the external magnetic field
and plays the role of the temperature in the classical Ising model.
Away from $\mu=\mu_c$, we conclude that $r_1(\mu,L)$ only depends on $L(\mu-1)$ in the critical regime,
 which is consistent with a correlation length diverging as $1/(\mu-1)$ close
to the critical point (located at $\mu_c=1$). In particular, we confirm the step-like shape of $r_1(\mu,L=\infty)$.

In Sec.~\ref{sec:n} we analyze the finite-size scaling  of $r_n(\mu=\mu_c=1,L)$ in the vicinity of $n=1/2$ and $n=1$, again with
the ICTF (up to $L=44$ sites).
The case $n=1/2$ turns out to be exactly solvable (Sec.~\ref{fig:n=0.5}) and related to
the ``ground-state degeneracy'' for a critical Ising model with free boundary conditions, as discussed by Affleck and Ludwig.\cite{al91}
In the vicinity of $n=1$ the numerical data strongly suggests
 a step-like shape of $r_n(\mu=1,L=\infty)$ as a function of the R\'enyi parameter $n$:
  $r_n(\mu=1,L=\infty)=0$ for $n<1$ and $r_n(\mu=1,L=\infty)=\ln 2$ for $n>1$.
This result has some important consequence regarding possible field theory approaches.
 In particular, a singularity at $n=1$ would invalidate any attempt
to compute $r_1$ from an analytical continuation 
to $n=1$ of the $n\in\mathbb N^*$ result (replica trick).

\section{Shannon entropy at the critical point}
\label{sec:2dTM}

\subsection{Square and triangular lattices}
\label{sec:sqandtri}
We compute the Shanon entropy $R_1$ using the transfer matrix $\mathcal T$ of the ferromagnetic Ising model.
We numerically diagonalize $\mathcal T$ (in the full space of dimension $2^L$), on the square and on the triangular lattices
\cite{foo4}
 for sizes up to $L=14$ and denote by $|L\rangle$ and $|E\rangle$ the left- and right- dominant eigenvectors of $\mathcal T$
 (corresponding to the eigenvalue with the largest modulus).
Then, the probability $p_i$
of a configuration $i$ is given by:
\begin{equation}
 p_i=\frac{\langle L|i\rangle\langle i|R\rangle}{\langle L|R\rangle}
\end{equation}
in the limit of a infinitely long cylinder $L_y \gg L$.

The results for $R_1(T_c)$, obtained by summing over the $2^L$ configurations, are shown in Fig.~\ref{fig:stc}.
The linear behavior, $R_1(T_c)\sim L$ is apparent, as well as the fact that the data for the two
lattices extrapolate to the same
value $\simeq  0.254$ at $L=0$. Although the systems are relatively small,
it shows that $r_1(T_c)\simeq  0.254$
does not depend on the microscopic lattice geometry, and is therefore very likely to be {\it universal}.

\subsection{Ising chain in transverse field}

As a third microscopic realization of the Ising 2d universality class, we study the ICTF:
\begin{equation}
 \mathcal{H}=-\mu \sum_{j=0}^{L-1} \sigma^x_j \sigma^x_{j+1} - \sum_{j=0}^{L-1}
\sigma^z_j.
	\label{eq:Hictf}
\end{equation}

{This Hamiltonian proportional to the logarithm of the transfer matrix of an anisotropic Ising model on the square lattice,
with couplings along the $y$ direction (``time'') which are much stronger than in the $x$ direction (``space'')
.

This Hamiltonian is transformed into a free fermion problem using the standard Jordan-Wigner
transformation. The later free fermion problem is then diagonalized using a Bogoliubov transformation.
 The ground-state of $\mathcal{H}$ is then described as the vacuum of the Bogoliubov fermions.
The critical point is located at $\mu=1$. For $\mu>1$ the system is in the ordered phase,
 with spontaneously broken $\mathbb Z_2$ symmetry
($\langle\sigma^x \rangle\ne 0$), and for $\mu<1$ the system is in the disordered (paramagnetic) phase.

It turns out that the ground-state $|G\rangle$ of the chain is simpler to express in $\sigma^z$ basis.
For an Ising spin configuration $|i\rangle$ labeled by the variables $\sigma_i^z=\pm 1$, the probability
at $\mu=1$ is :
\begin{equation}
 p_i=|\langle i | G\rangle|^2=p(\sigma_0^z,\cdots,\sigma_{L-1}^z)=\det M
\label{eq:pidet}
\end{equation}
where $M$ is an $L\times L$ matrix defined by:
\begin{equation}
 M_{j\ell} = \frac{1}{2}\delta_{j\ell}+
\frac{(-1)^{j-\ell}\sigma_j^z}{2L\sin\left[\pi(j-\ell+\frac{1}{2})/L\right]}
\label{eq:Mij}
\end{equation}
This result is derived in Appendix.~\ref{sec:ICTF_diag}, where the non-critical case $\mu \neq 1$ is also considered. 
However, going back to the initial 2d classical model,
the actual  spin directions are measured by $\sigma^x_i$.
So, we first compute
an entropy $R^{(z)}_n$ corresponding to probabilities of $z$-axis configurations, and then use
the Kramers-Wannier duality transformation\cite{kw41} to 
obtain the desired $R_n=R^{(x)}_n$: 
\begin{equation}
 R^{(x)}_n (\mu)=R^{(z)}_n (1/\mu)+\ln 2.
\label{eq:sxsz}
\end{equation}

The calculation of $R_n^z$ amounts to compute $2^L$ probabilities, each of which is obtained as a determinant of size $L\times L$.
Using the translation invariance and the reflection symmetry of the chain, the number of probabilities to compute
can be reduced to  $\sim 2^L/(2L)$.
\cite{foo5}
To do so we generate one representative for each orbit of spin configurations
 (under the action of the lattice symmetries) using  the ``bracelets'' enumeration
 algorithm of Ref.~\onlinecite{sawada01}.
For the largest size, $L=44$ , computing all the probabilities ($2^L=1.7\times10^{13}$) required about
 one thousand hours of CPU time on a parallel machine.

The data for the Shannon entropy $R_1$ are plotted in Fig.~\ref{fig:stc} and given in Table \ref{tab:r1}.
They significantly extend the results published in Ref.~\onlinecite{stephan09}.
The columns $r_1^{(1)}$, $r_1^{(4)}$ and $r_1^{(5)}$  correspond to three different ways
 to extract the subleading constant from $R_1(\mu=1,L)$, with three different types of fits
 (details in the table caption). In all cases the result rapidly converges and, 
using the largest size ($L=44$ spins)
we estimate that $r_1=0.2543925(5)$ at $L=\infty$.

\begin{figure}
\begin{center}
\includegraphics[width=8cm]{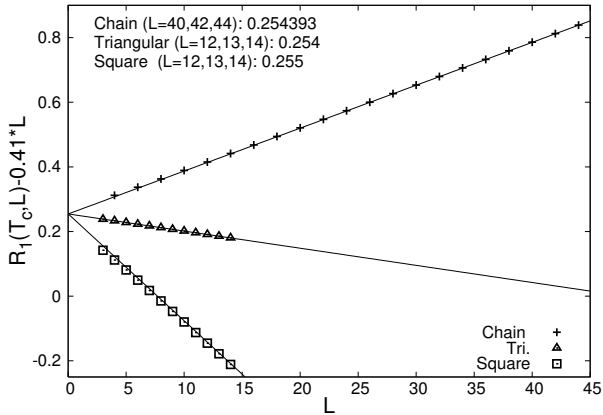}
\caption[1]{Shannon entropy $R_1(T_c,L)$ of the Ising models at the critical point, plotted
as a function of $L$ 
(a linear term, $-0.41 L$, has been subtracted for clarity).
Data for the square and triangular lattices and for the Ising chain in transverse field
are compared. 
The data
are well reproduced by $R_1(L)\simeq aL+r_1+b/L$
and the subleading constant $r_1$ is evaluated using the three largest sizes.
Each line represents the leading term  and the constant, $aL+r_1$.
The subleading term appears to be the same
$r_1 \simeq 0.254$ for the three microscopic models.
} 
\label{fig:stc}
\end{center}
\end{figure}

\begin{table*}
 \begin{tabular}{|c|c|c|ccc|}
\hline \\
$L$ & $R_1(L)$ & $a_1$ & $r_1^{(1)}$ & $r_1^{(4)}$& $r_1^{(5)}$ \\
\hline
16& 7.02789845748593     & 0.4232735600  & \textcolor{red}{0.254}4012149 & \textcolor{red}{0.254392}4985  & \textcolor{red}{0.2543925}471\\
18& 7.87432026832476     & 0.4232735603  & \textcolor{red}{0.25439}83072 & \textcolor{red}{0.2543925}177  & \textcolor{red}{0.2543925}302\\
20& 8.72076710746883     & 0.4232735604  & \textcolor{red}{0.25439}65648 & \textcolor{red}{0.2543925}180  & \textcolor{red}{0.2543925}183\\
22& 9.56723215961776     & 0.4232735605  & \textcolor{red}{0.25439}54570 & \textcolor{red}{0.2543925}156  & \textcolor{red}{0.2543925}130\\
24& 10.4137108773778    & 0.4232735605  & \textcolor{red}{0.25439}47190  & \textcolor{red}{0.2543925}136  & \textcolor{red}{0.2543925}110\\
26& 11.2602001105626    & 0.4232735606  & \textcolor{red}{0.25439}42083  & \textcolor{red}{0.2543925}110  & \textcolor{red}{0.2543925}072\\
28& 12.1066976079502    & 0.4232735605  & \textcolor{red}{0.25439}38437  & \textcolor{red}{0.2543925}139  & \textcolor{red}{0.2543925}188\\
30& 12.9532017180203    & 0.4232735608  & \textcolor{red}{0.25439}35763  & \textcolor{red}{0.2543925}001  & \textcolor{red}{0.2543924}741\\
32& 13.7997112017585    & 0.4232735600  & \textcolor{red}{0.25439}33760  & \textcolor{red}{0.2543925}306  & \textcolor{red}{0.2543925}939\\
34& 14.6462251114521    & 0.4232735614  & \textcolor{red}{0.25439}32227  & \textcolor{red}{0.2543924}796  & \textcolor{red}{0.2543923}635\\
36& 15.4927427098430    & 0.4232735597  & \textcolor{red}{0.25439}31036  & \textcolor{red}{0.2543925}326  & \textcolor{red}{0.2543926}640\\
38& 16.3392634147881    & 0.4232735610  & \textcolor{red}{0.25439}30095  & \textcolor{red}{0.2543925}037  & \textcolor{red}{0.2543924}262\\
40& 17.1857867605076    & 0.4232735606  & \textcolor{red}{0.254392}9343  & \textcolor{red}{0.2543924}999  & \textcolor{red}{0.2543924}890\\
42& 18.0323123698967    & 0.4232735603  & \textcolor{red}{0.254392}8734  & \textcolor{red}{0.2543925}161  & \textcolor{red}{0.2543925}658\\
44& 18.8788399343835    & 0.4232735602  & \textcolor{red}{0.254392}8237  & \textcolor{red}{0.2543925}300  & \textcolor{red}{0.2543925}757\\
\hline
\end{tabular}
\caption{(Color online)
Shannon entropy $R_1(L,\mu=1)$ of the critical Ising chain in transverse field as a function of the system size $L$.
The subleading constant $r_1$ is extracted using three different fits:
$r_1^{(1)}$ is obtained by a fit
to $R_1(L)\simeq aL+r_1+bL^{-1}$ using the three following system sizes: $L,L-2,L-4$.
$r_1^{(4)}$ is obtained by a fit
to $R_1(L)\simeq aL+r_1+bL^{-1}+\cdots eL^{-4}$ using the six system sizes $L,L-2,\cdots,L-10$.
$r_1^{(5)}$ is obtained by a fit
to $R_1(L)\simeq aL+r_1+bL^{-1}+\cdots f L^{-5}$ using the seven system sizes $L,L-2,\cdots,L-12$.
$a$ is the coefficient of the extensive (and non-universal) term, extracted from the seven-point fit above.
From this analysis, our best estimate for $L=\infty$ is
$r_1=0.2543925(5)$.
}\label{tab:r1}
\end{table*}

\section{$\mu$ away from the critical point}
\label{sec:mu}
In this section, we investigate the behavior of $r_1$ in the vicinity of the critical point, by
considering the Ising chain in transverse field away from $\mu=1$.
The results are summarized in Fig.~\ref{fig:mu}.

In this plot, $r_1(\mu)$ is extracted from $R_1(L,\mu)$ using a fit to $a_1(\mu) L+r_1(\mu)+b_1(\mu)/L$ with three
consecutive values of $L$.
For the size we have studied (here $L\leq 38$), there is still some visible finite-size effects.
In particular, the marked oscillations 
in the vicinity of $\mu=1$ are not converged to the $L=\infty$ limit.
In fact, 
it is reasonable to  expect the curves to gradually approach a step-like function as $L$ increases: $r_1=0$ for $\mu<1$
and $r_1=\ln(2)$ for $\mu>1$.

This scenario, anticipated in Ref.~\onlinecite{stephan09}, is corroborated by the scaling shown in the inset of Fig.~\ref{fig:mu}. When plotted as a function of
$(\mu-1)L$, the data for different system sizes and different values of $\mu$ collapse
onto a single -- and very likely universal -- curve.
 This can be understood from the fact that the correlation length 
$\xi$ of the Ising model diverges as $1/| \mu-1|$ at the transition, and if one assumes that
$r_1$ is a function of $L/\xi(\mu)$ in the critical region. If correct, it immediately implies
that $r_1(\mu)$ is a step-like function in the thermodynamic limit.

\begin{figure}
\begin{center}
\includegraphics[width=6cm,angle=-90]{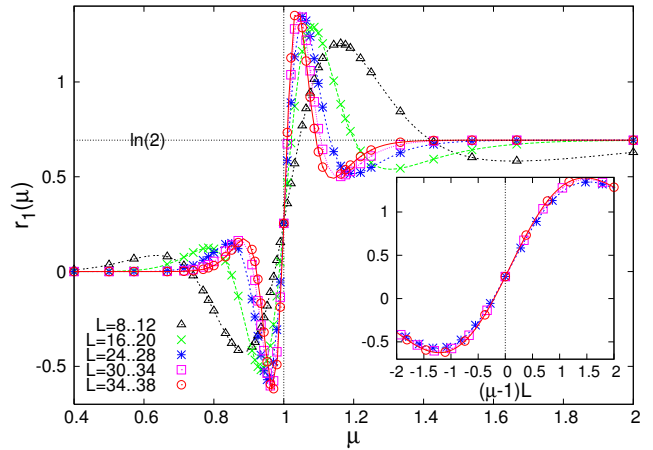}
\caption[1]{Subleading constant $r_1$ of the Shannon entropy of the Ising chain as a function of $\mu$.
The critical
point corresponds to $\mu=1$.
Inset:  $r_1(\mu)$ for different system sizes, plotted as a function of $(\mu-1)L$.
}\label{fig:mu}
\end{center}
\end{figure}

\section{R\'enyi entropy away from $n=1$}
\label{sec:n}

We now consider the effect of changing the R\'enyi parameter $n$.
When $2n$ is an integer, $R_n$
has an interpretation in terms of the free energy of $k=2n$ semi-infinite Ising models which are ``glued'' together at their boundary (see Fig.~\ref{fig:gluing}).
\begin{figure}
 \begin{center}
  \includegraphics[width=8cm]{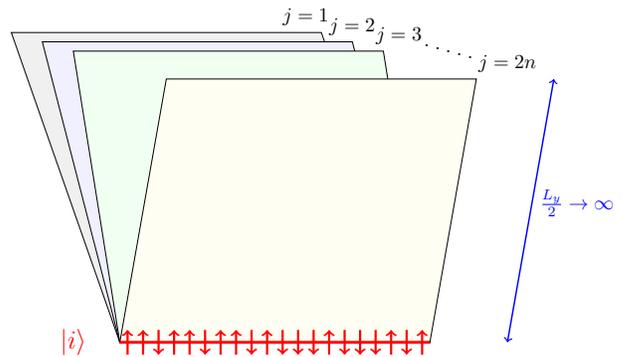}
\caption[1]{$2n$ Ising models glued together at their boundary (``Ising book'').
 In our case, each ``page'' has periodic boundary conditions along the horizontal axis and is semi infinite in the vertical direction.
 Fig.~\ref{fig:cyl} corresponds to $2$ pages ($n=1$).
}
\label{fig:gluing}
 \end{center}

\end{figure}

Using the transfer matrix point of view,
it is simple to see that  $p_i^{k/2}$ is  (proportional to) the probability
 to observe the spin configuration $i$ on a circle along which $k$ Ising models
 (defined on semi-infinite cylinders) are forced to coincide.
This  was used in Refs.~\onlinecite{fm06} and \onlinecite{hsu09} in some field theory calculations, but
it is also true at the microscopic level.
The interpretation above does not apply when $2n$ is not a positive integer, but
$R_n(L,\mu)$ can still be computed numerically for any $n\geq0$.

\begin{figure*}
\begin{center}
\includegraphics[width=11.5cm,angle=-90]{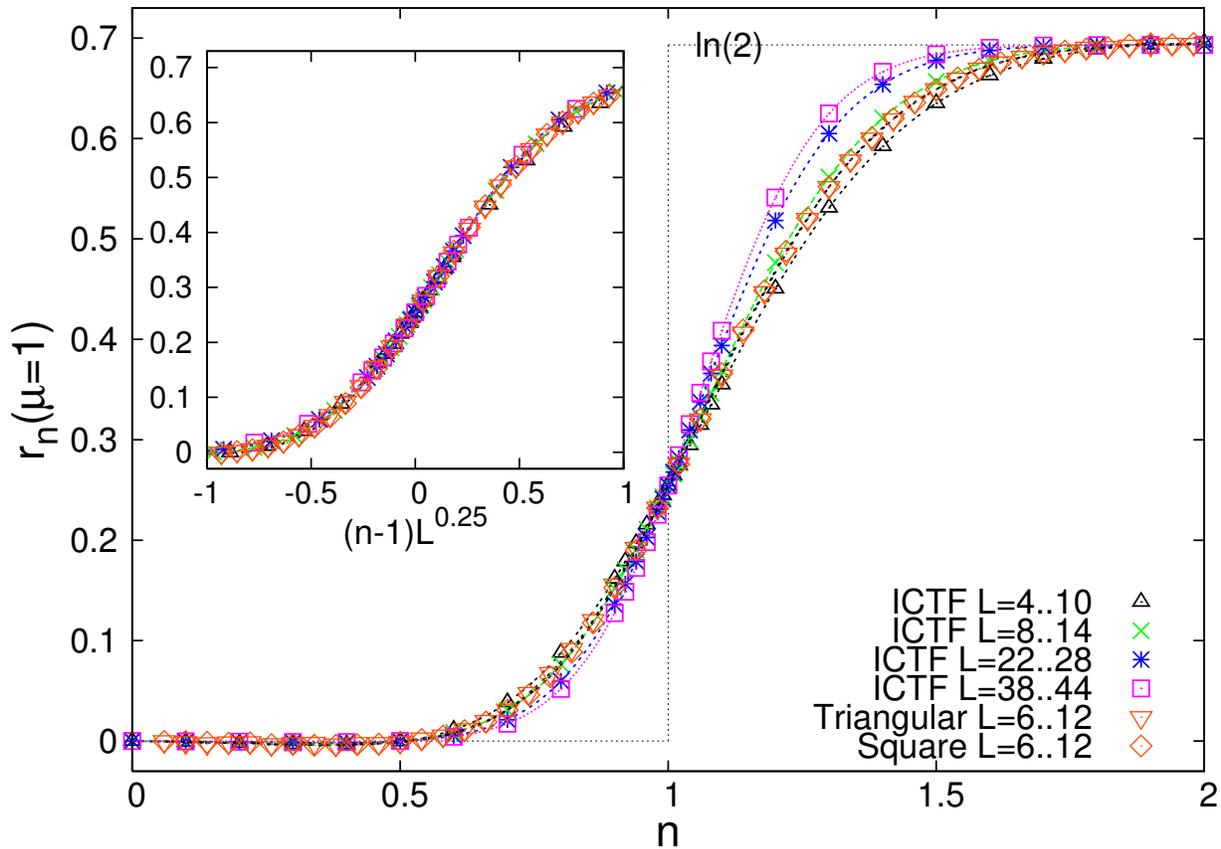}
\caption[1]{(color online) Subleading constant $r_n$ of the R\'enyi entropy
 of the Ising chain (ICTF, at $\mu=1$) and classical Ising models (triangular and
 square lattice, at $T=T_c$) as a function of the Renyi parameter $n$.
The (slow) convergence towards a step function can be observed.
 Inset : when plotted as a function of $(n-1)L^{0.25}$, the data collapse reasonably well onto a single curve.
 For each value of $n$, $r_n$ is obtained by fitting the data for $R_n(L)$ to $\simeq aL+r_n +bL^{-1}+cL^{-2}$ using
 four system size : $L,L-2,L-4$ and $L-6$, as indicated.
}\label{fig:renyi}
\end{center}
\end{figure*}

\subsection{R\'enyi parameter $n=2$ and above}
\label{sec:above2}

When $n$ goes to infinity, only the spin configuration with the largest
probability contributes to $R_n$. For the ferromagnetic Ising models we consider (including
the quantum chain in transverse field), this configuration is two-fold degenerate and
corresponds to a fully polarized ferromagnetic state, $|\!\uparrow\uparrow\cdots\uparrow\rangle$
or $|\!\downarrow\downarrow\cdots\downarrow\rangle$. In other words,
taking the limit $n\to\infty$ amounts to study a semi-infinite Ising model with ferromagnetic boundary conditions.
The corresponding probability, $p_{\rm max}$, behaves as $-\ln(p_{\rm max})\sim aL+\ln(2)$ at the critical point.\cite{stephan09}
The subleading constant, $\ln(2)$, is nothing but (twice) the ``$g$ factor'' associated to this conformally invariant boundary condition
(more details in Sec.~\ref{sec:n=0.5}).
This implies
for the R\'enyi entropies that the subleading constant $r_n(\mu_c)$ is $\ln(2)$ at $n=\infty$.

In fact, for $n\gtrsim2$ the Fig.~\ref{fig:renyi} shows that even relatively small
systems give $r_n(\mu_c)$ very close to $\ln 2$.
Table \ref{tab:r2} is an  analysis showing that $r_2(\mu=1)\simeq\ln 2$ with a great accuracy,
of the order of $10^{-8}$.
Since the convergence to $\ln 2$ is even faster when $n>2$, there
is practically no doubt that $r_n(\mu_c)$ is exactly $\ln 2$ for $n\gtrsim 2$.
\cite{foo6}

As a consequence, an analytical continuation of this result to $n=1$ would erroneously give $r_1(\mu_c)=\ln 2$ (instead of 0.25439).
In particular, we note that the results of Ref.~\onlinecite{hsu09} (which use a replica technique)
 are in agreement with ours for $n>1$, but not {\it at} $n=1$.

\begin{table*}
 \begin{tabular}{|c|c|c|ccc|}
\hline \\
$L$ & $R_2(L)$ & $a_2$ & $r_2^{(1)}/\ln 2$ & $r_2^{(4)}/\ln 2$& $r_2^{(5)}/\ln 2$ \\
\hline
20      & 4.95205232373074     & 0.2138075040  & 0.9989748222  & 0.9999928713  & 0.9999877126\\
28      & 6.66741530818944     & 0.2138074244  & 0.9996525352  & 0.9999971726  & 0.9999989449\\
36      & 8.38060934985332     & 0.2138074200  & 0.9998432968  & 0.9999991645  & 0.9999998996\\
44      &10.0928119559937    & 0.2138074203  & 0.9999165184  & 0.9999996643  & 0.9999997718\\
\hline
\end{tabular}
\caption{R\'enyi entropy $R_2(L,\mu=1)$ of the critical Ising chain in transverse field as a function of the system size $L$.
The subleading constant $r_2(\mu_c)$ is extracted using three different fits
(same as in Table \ref{tab:r1}):
$r_2^{(1)}$ is obtained by a fit
to $R_2(L)\simeq aL+r_1+bL^{-1}$ using the three following system sizes: $L,L-2,L-4$.
$r_2^{(4)}$ is obtained by a fit
to $R_2(L)\simeq aL+r_1+bL^{-1}+\cdots eL^{-4}$ using the six system sizes $L,L-2,\cdots,L-10$.
$r_2^{(5)}$ is obtained by a fit
to $R_2(L)\simeq aL+r_1+bL^{-1}+\cdots f L^{-5}$ using the seven system sizes $L,L-2,\cdots,L-12$.
$a$ is the coefficient of the extensive (and non-universal) term, extracted from the seven-point fit above.
These data show that  $r_1$ converges to $\ln 2$ (limit $L\to \infty$).
Similar results, with an even faster convergence, show that $r_{n\geq2}(\mu_c)=\ln(2)$. With the present
 systems sizes and the present machine accuracy, adding more terms
in the 1/L expansion does not increase the accuracy on $r_2$.
}\label{tab:r2}
\end{table*}

\subsection{$n=\frac{1}{2}$}
\label{sec:n=0.5}

The special value $n=\frac{1}{2}$ corresponds
to the free energy of a {\it single} Ising model defined on a semi-infinite cylinder (keeping only part $A$ in Fig.~\ref{fig:cyl}), and
can be treated exactly.
Using the transfer matrix point of view, it is indeed simple to see that $\sqrt{p_i}$ is proportional
to the probability to observe the spin configuration $i$
at the edge of a semi-infinite Ising model (contrary to $p_i$ which is the probability to observe $i$ in the {\it bulk}).

As far as the universal properties are concerned, we can study the ground state
of the quantum Ising chain (Eq.~(\ref{eq:Hictf})) rather than the transfer matrix of the classical 2d model.
Denoting by $|G\rangle$ the ground-state of the chain, we have $\sqrt{p_i}=\langle G | i\rangle$
and the R\'enyi entropy $R_\frac{1}{2}$ can be written as:
\begin{eqnarray}
 R_{\frac{1}{2}}(L)&=&2 \ln \left( \langle G |\sum_{\{\sigma^x_i=\pm1\}} | \sigma^x_1 \cdots \sigma^x_L\rangle \right) \\
	&=&2 \ln \Big(2^{L/2}\langle G | \rm free\rangle\Big) \label{eq:R05}
\end{eqnarray}
where $| {\rm free}\rangle=|\{\sigma^z_i=1\}\rangle$ is the state where all the spins point in the $z$ direction.
It turns out that the latter state is the vacuum of Jordan-Wigner Fermion and that the scalar product
in Eq.~(\ref{eq:R05}) can be obtained as a particular case of Eqs.~(\ref{eq:pidet}-\ref{eq:Mij}).
At the critical point ($\mu=1$), the result is particularly simple\cite{stephan09}
\begin{equation}
 \langle G |{\rm free}\rangle=\prod_{j=0}^{L/2-1} \cos \frac{(2j+1)\pi}{4L},
\end{equation}
and leads to the following exact expression of the $n=\frac{1}{2}$-R\'enyi entropy:
\begin{equation}
 R_{\frac{1}{2}}(L,\mu=1)=L\ln 2+2\sum_{j=0}^{L/2-1} \ln \cos\frac{(2j+1)\pi}{4L}.
\end{equation}
Finally, an Euler-Maclaurin expansion 
gives the desired finite-size scaling,
with a vanishing constant $r_{\frac{1}{2}}$:
\begin{eqnarray}
 R_{\frac{1}{2}}(L,\mu=1)&=&a_{\frac{1}{2}} L +r_{\frac{1}{2}}+o(1) \\
	a_{\frac{1}{2}}&=&\frac{2K}{\pi} \\
	r_{\frac{1}{2}}&=&0
\label{eq:Kpi}
\end{eqnarray}
where $K\simeq 0.91596559$ is Catalan's constant.

The constant term in $-\ln \langle G | {\rm free}\rangle$
has already been studied in Ref.~\onlinecite{stephan09}. The situation where $|G\rangle$
is the ground-state of an antiferromagnetic spin-$\frac{1}{2}$ XXZ chain has also been considered.\cite{cvsz09,stephan09}
Such a scalar product is closely related to the notion of quantum fidelity.\cite{cvsz09}
In terms of a classical 2d Ising model, $-T \ln \langle G | {\rm free}\rangle$ is the boundary contribution
 to the free energy of a semi-infinite Ising model with free boundary conditions imposed at the edge.
 At the critical point, this  is a well understood quantity from boundary CFT,
and the subleading constant $r_{\frac{1}{2}}$ corresponds
to $-2\ln g$, where $g$ is the ``ground-state degeneracy'' discussed by Affleck and Ludwig.\cite{al91}
In the present case of the Ising model, $r_1=0$ is in agreement with $g_{\rm free}=1$.\cite{al91,cardy89}
 This result has also been checked numerically in Ref.~\onlinecite{djs09}

\subsection{Critical behavior in the vicinity of $n=1$}
\label{sec:n_close_to_1}
The results concerning the subleading constant $r_n(\mu=1)$ are summarized in Fig.~\ref{fig:renyi}.
The behavior of $r_n(\mu=1)$ has some similarity with
that of $r_1(\mu)$: the curves interpolates between 0 and $\ln 2$ with a slope
at $n=1$ (resp. $\mu=1$) which increases as a function of the system size.
Here again, it appears that the data for different values of $n$ and $L$ collapse
onto a single curve when plotted as a function of $(n-1)L^{0.25}$ (inset of  Fig.~\ref{fig:renyi}).
The error bar on the exponent 0.25 are unfortunately large and difficult to estimate,
but it indicates (a rather slow) divergence of the slope $\partial r_n /\partial n|_{n=1}$ when $L$ increases.
 $r_n$ has also been computed for the classical Ising models on the square and triangular lattices, as in
 Sec.~\ref{sec:sqandtri}. The inset of Fig.~\ref{fig:renyi} shows that the $r_n$ obtained from the corresponding transfer matrix
 calculations are in good agreement with those calculated from the ground state of the ICTF. This is a strong indication that,
 in a scaling region around $n=1$, $r_n$ defines a {\it universal curve}.
The analogy between the effects of $\mu$ and $n$ suggests that $n-1$ is a ``relevant'' perturbation:
going slightly below (resp. above) $n=1$ induces a drastic change in $r_n(\mu=1)$, which immediately (when $L=\infty$)
goes to $0$ (resp. $\ln 2$), as in high (resp. low) temperature phase of the 2d Ising model.

\subsection{Vicinity of $n=\frac{1}{2}$}

The value $n=\frac{1}{2}$ can be treated exactly, as explained in Sec.~\ref{sec:n=0.5}.
However, the free-fermion calculation does not extend away from $n=\frac{1}{2}$.
Still, at $n=0.5$ we observe (numerically) a crossing of the curves corresponding to different values of $L$ (see Fig.~\ref{fig:n=0.5}). 
This phenomenon, also observed at $n=1$, is reminiscent of a critical behavior, where the deviation
 away from $n=1/2$ would play the role of an irrelevant perturbation away from a fixed point.
The data can also be collapsed onto a single curve, when the y-axis is multiplied by a factor $\approx L^{0.6}$.
 However, contrary to the case $n=1$, this result
 indicates a reasonably fast convergence towards $r_n=0$ in the vicinity of $n=1/2$ (see the inset of Fig.~\ref{fig:n=0.5}).

\begin{figure}
\begin{center}
\includegraphics[width=6cm,angle=-90]{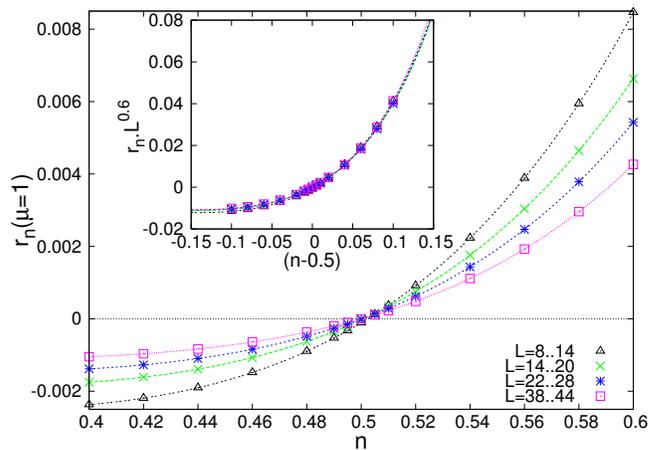}
\caption[1]{(color online) Subleading constant $r_n$ of the R\'enyi entropy of the Ising chain (at the critical point $\mu=1$),
in the vicinity of $n=0.5$. Inset : subleading constant multiplied by $L^{0.6}$ as a function of $(n-0.5)$.
}\label{fig:n=0.5}
\end{center}
\end{figure}

\section{Discussion and conclusions}
In the present Ising models, $r_n(\mu)$ seems to take only three discrete values. 
For example in the critical case, we find :
\begin{equation}
r_n(\mu=1)=\left\{ \begin{array}{ccc}
0&,&n<1\\
0.2543925(5) &,&n=1\\
\ln 2 &,& n>1
\end{array}\right..
\end{equation}

This is quite different from other models
described in terms of a free field compactified (with radius $R$) in the long-distance limit.
In that case, which is better understood from a field theory point of view, the system
describes a line of fixed points and the subleading constant $r_n(R)$ continuously varies along that critical line\cite{foo7}
\begin{equation}
 r_n(R)=\ln R -\frac{\ln n}{2(n-1)}.
\end{equation}

We have discussed how the special values $n=\frac{1}{2}$, and $n=\infty$ are related to the $g$ factors associated
to free and fixed boundary conditions of
the Ising model. But so far, we do not know how to understand $r_{n=1}$ for the critical Ising model using CFT. This is certainly an interesting 
question for future studies. This is all the more challenging, as it seems (Secs.~\ref{sec:above2} and \ref{sec:n_close_to_1}) that replica
methods for the R\'enyi parameter $n$ are not applicable to the Ising critical point for this quantity.

It is tempting to conjecture that crossings for $r_n(L)$ are observed whenever the underlying probabilities, $\sim p_i(\mu)^n$
 describe a conformally invariant setup. It is indeed the case at $n=1/2$ (Ising boundary with free boundary conditions),
 but it is also realized for $n=1$, since it correspond to the bulk probabilities.
 It would be interesting to check this idea on other models, and
to investigate the likely connection with the theory of line defects in conformally invariant systems.\cite{oa96,pz01}.

{\it Acknowledgments --- } 
We wish to thank J. Dubail, S. Furukawa, A. La\"uchli, Ph. Lecheminant, M. Oshikawa and H. Saleur for several useful discussions and suggestions.

The numerical calculations were done on the machine {\tt titane} at the ``Centre de calcul centralis\'e du CEA''
under the project number p575.
\appendix
\section{Probability of a spin configuration}
\label{sec:ICTF_diag}
We consider an Ising chain in transverse field 
\begin{equation}
 \mathcal{H}=-\mu \sum_{j=0}^{L-1} \sigma_j^x \sigma_{j+1}^x -\sum_{j=0}^{L-1} \sigma_j^z.
\end{equation}
We assume $L$ to be even, as well as periodic boundary conditions $\sigma_L^x=\sigma_0^x$. We wish to find the ground state $|G\rangle$ of
 this Hamiltonian, and to compute all of his components in the basis of the eigenstates of the $\sigma_j^z$.
\subsection{Diagonalization}
As is well known, $\mathcal{H}$ can be expressed in terms of free fermions, using the Jordan-Wigner transformation :
\begin{eqnarray}
 \sigma_j^x+i\sigma_j^y&=&2c_j^\dag \exp \left(i\pi \sum_{l=0}^{j-1}c_l^\dag c_l\right)\\
\sigma_j^z&=&2c_j^\dag c_j -1,
\end{eqnarray}
where the $c,c^\dag$ satisfy the canonical anticommutation relations $\{c_j,c_\ell^\dag\}=\delta_{j\ell}$.
 This allows to write the Hamiltonian as a quadratic form
\begin{equation}
 \mathcal{H}=-\sum_{j=0}^{L-1}(2c_j^\dag c_j -1)-\mu \sum_{j=0}^{L-1}(c_j^\dag -c_j)(c_{j+1}^\dag +c_{j+1}),
\end{equation}
where the fermions are subject to the following boundary condition :
\begin{equation}
 c_{L}^\dag=-\exp\left(i\pi \mathcal{N}\right)c_0^\dag \quad,\quad \mathcal{N}= \sum_{l=0}^{L-1}c_l^\dag c_l.
\end{equation}
The parity operator $\mathcal{P}$ commutes with $\mathcal{H}$
\begin{equation}
 \mathcal{P}=\prod_{j=0}^{L-1}\sigma_j^z=\exp\left(i\pi\mathcal{N}\right)=\pm 1,
\end{equation}
and because of Perron-Frobenius theorem, the ground-state lies in the sector $\mathcal{P}=+1$. Therefore, 
fermions are subjected to antiperiodic boundary conditions $c_L^\dag=-c_0^\dag$.
 $\mathcal{H}$ can finally be diagonalized by a Bogoliubov transformation :
\begin{eqnarray}\label{eq:bogo}
 c_j^\dag &=&\frac{1}{\sqrt{L}}\sum_{k}e^{ikj}\left(\cos \theta_k d_k-i \sin \theta_k d_{-k}^\dag\right)\\
k&\in& \left\{(2l+1)\pi/L\,\big|-L/2\leq l \leq L/2-1\right\}\\
&\sin& 2\theta_k=\frac{\mu \sin k}{\sqrt{1+2\mu \cos k+\mu^2}}\\
&\cos& 2\theta_k=\frac{1+\mu \cos k}{\sqrt{1+2\mu \cos k+\mu^2}}
\end{eqnarray}
The new fermions operators $d_k,d_k^\dag$ satisfy the necessary anticommutation relations, and diagonalize $\mathcal{H}$ :
\begin{eqnarray}
 \mathcal{H}&=&\sum_k \varepsilon_k \left(d_k^\dag d_k -1/2\right)\\
\varepsilon_k&=&2\sqrt{1+2\mu \cos k+\mu^2}.
\end{eqnarray}
$\varepsilon_k>0$ ensures that the ground-state $|G\rangle$ is the vacuum $|0\rangle$ of the $d_k$.
\subsection{Exact formulae for the spin probabilities}
We define $P_j^\sigma$ as the projector onto the $|\sigma\!=\!\pm 1\rangle_j^z$ state:
\begin{equation}
 P_j^{+}=c_j^\dag c_j\quad,\quad P_j^- =c_j c_j^\dag. 
\end{equation}
$p_i$ is then given by
\begin{equation}
 p_i=p(\sigma_0,\ldots,\sigma_{L-1})=\langle 0 |P_0^\sigma P_1^\sigma \ldots P_{L-1}^\sigma|0\rangle.
\end{equation}
Using Wick's theorem, this correlator reduces to a Pfaffian. To compute it, we need to calculate the four types of contractions $\langle c_j^\dag c_\ell\rangle$,
 $\langle c_j c_\ell^\dag\rangle$, $\langle c_j^\dag c_\ell^\dag\rangle$, $\langle c_j c_\ell\rangle$, which can be done using Eq.~(\ref{eq:bogo}).
 It is worth noticing that all these correlators are real in this particular model. We write a generic projector as :
\begin{equation}
 P_j^\sigma=f_{j}^\dag f_{j},
\end{equation}
with $f_j^\dag=c_j^\dag$ for $\sigma=+1$ and $f_j^\dag=c_j$ for $\sigma=-1$. Then :
\begin{eqnarray}
 p_i^2&=&\big\langle f_0^\dag f_0 f_1^\dag f_1\ldots\ldots f_{L-1}^\dag f_{L-1}\big\rangle^2\\ 
&=&\big\langle f_0^\dag f_1^\dag \ldots f_{L-1}^\dag f_0 f_1 \ldots f_{L-1}\big\rangle^2\\\label{eq:pfaff}
&=&\textrm{Pf}^{\,2}\,\left(\begin{array}{cc}A&B\\-B&-A\end{array}\right)
\end{eqnarray}
where $A$ is antisymmetric, $B$ is symmetric, and $\textrm{Pf}$ denotes the Pfaffian.
 The matrix elements of $A$ and $B$ are 
\begin{eqnarray}
 A_{j\ell}&=&\langle f_j^\dag f_\ell^\dag\rangle\quad,\quad \ell\geq j\\
B_{j\ell}&=&\langle f_j^\dag f_\ell\rangle
\end{eqnarray}
Using the relation $\textrm{Pf}^{\,2}=\det$, Eq.~(\ref{eq:pfaff}) simplifies into
\begin{eqnarray}
\label{eq:proba1}
p_i^2&=&\det \left(\begin{array}{cc}A&B\\-B&-A\end{array}\right)\\\label{eq:proba2}
&=&\det \left(\begin{array}{cc}A+B&B\\0&B-A\end{array}\right).
\end{eqnarray}
Eq.~(\ref{eq:proba2}) follows from Eq.~(\ref{eq:proba1}) by adding the second column to the first,
 and then the first row to the second. Finally, 
\begin{equation}
 p_i=\det (A+B)=\det M,
\end{equation}
 where $M$ is a $L\times L$ matrix with elements
\begin{eqnarray}
 M_{j\ell}&=&\langle f_j^\dag (f_\ell^\dag + f_\ell)\rangle\\
&=&\frac{1}{2}\delta_{j \ell}+\frac{\sigma_j^z}{2L}\sum_k \cos [k(j-\ell)+2\theta_k]
\end{eqnarray}
At the critical point ($\mu=1$), $\theta_k=k/4$ and the matrix elements simplify even further :
\begin{equation}
 M_{j\ell}=\frac{1}{2}\delta_{j \ell}+\frac{ (-1)^{j-\ell}\sigma_j^z}{2L\sin [\pi(j-\ell+1/2)/L]}.
\end{equation}


\begin{thebibliography}{99}
\bibitem{foo1}
This quantum/classical correspondence works in a rather straightfoward way
 for simple constrained models (such as dimer models or vertex models).
 For other models, such as the Ising model considered in this paper,
 some additional care is needed to define the geometry of the $A/B$ boundary at the microscopic level.
 In the particular case of 2d classical Ising models, the spins living at the frontier between $A$ and $B$ have
 to be ``duplicated'' to insure that the decomposition induced by the
 classical spin configurations is indeed a proper Schmidt decompostion
of the RK state. See Ref.~\cite{stephan09} for more details.
\bibitem{foo2}
In the quantum point of view,
 where one studies the entanglement in a RK wave-function, the dominant ($\sim L$)
 contribution is the boundary (also called ``area'') law.
\bibitem{foo5}
We also use the property that, for periodic boundary conditions,
 $\prod_i \sigma^z_i=1$ in the ground state $|G\rangle$.
 Since we work in the $\sigma^z$ basis, this reduces by another factor two the number of probabilities to compute.
\bibitem{foo6}
In fact, the analysis of Sec.~\ref{sec:n_close_to_1}
 suggests that $r_n(\mu_c)$ flows to its ferromagnetic boundary condition limit, $\ln(2)$, as soon as $n>1$.
\bibitem{foo7}
See Eq.~(77) in Ref.~\onlinecite{stephan09}. The following normalization of the radius is used: $R=1$ for free fermions, and $R=\sqrt 2$ at $SU(2)$
 symmetric (self-dual) point of the XXZ chain. This corresponds to a lagrangian $\mathcal{L}=\frac{1}{8\pi} \left(\partial_\mu \varphi\right)^2$,
 where the field is compactified on a circle of radius $R$ : $\varphi=\varphi+2\pi R$.
\end{thebibliography}
\end{document}